\documentclass[amstex,amsfonts,aps,prb,overcite,twocolumn,amsmath]{revtex4}
\usepackage{graphicx}

\begin{document}
\title{\textbf{An Empirical Approach to the Bond Additivity Model in Quantitative Interpretation of Sum Frequency Generation Vibrational Spectra}}
\author{Hui Wu$^{a}$\footnote[2]{These two authors are with equal
contribution.}}
\author{Wen-kai Zhang$^{b}$\footnotemark[2]}
\author{Wei Gan$^{b}$}
\author{Zhi-feng Cui$^{a}$}
\author{Hong-fei Wang$^{a,b}$\footnote[1]{Authors to whom correspondence should be addressed.
E-mail: hongfei@mrdlab.icas.ac.cn, Tel: 86-10-62555347, Fax:
86-10-62563167.}} \affiliation{a. Department of Physics, Anhui
Normal University, Wuhu, Anhui Province, China 241000; \\ b. State
Key Laboratory of Molecular Reaction Dynamics,
\\Institute of Chemistry, the Chinese Academy of Sciences,
Beijing, China, 100080}
\date{\today}

\begin{abstract}
Knowledge of the ratios between different polarizability $\beta
_{i'j'k'}$ tensor elements of a chemical group in a molecule is
crucial for quantitative interpretation and polarization analysis of
its SFG-VS spectrum at interfaces. The bond additivity model or the
hyperpolarizability derivative model along with experimentally
obtained Raman depolarization ratios has been widely used to obtain
such tensor ratios for the CH$_{3}$, CH$_{2}$, and CH groups.
Successfully, such treatment can quantitatively reproduce the
intensity polarization dependence in SFG-VS spectra for the
symmetric (ss) and asymmetric (as) stretching modes of CH$_{3}$ and
CH$_{2}$ groups, respectively. However, the relative intensity
between the ss and as modes usually does not agree with each other
within this model even for some of the simplest molecular systems,
such as the air/methanol interface. This fact certainly has cast
uncertainties on the effectiveness and conclusions based on the bond
additivity model. One of such examples is that the as mode of
CH$_{3}$ group has never been observed in SFG-VS spectra from the
air/methanol interface, while this as mode is usually very strong
for SFG-VS spectra from the air/ethanol, other short chain alcohol,
as well as long chain surfactants, interfaces. In order to answer
these questions, an empirical approach from known Raman and IR
spectra is used to make corrections to the bond additivity model.
With the corrected ratios between the $\beta _{i'j'k'}$ tensor
elements of the ss and as modes, all features in the SFG-VS spectra
of the air/methanol and air/ethanol interface can be quantitatively
interpreted within the bond additivity model. This empirical
approach not only provides new understandings of the effectiveness
and limitations of the bond additivity model, but also provides a
practical roadmap for its application in SFG-VS studies of molecular
interfaces.
\end{abstract} \pacs{42.65.Ky, 68.18.Jk}

\maketitle
\section{Introduction}
Since the first report of Sum Frequency generation Vibrational
Spectroscopy (SFG-VS) experiment in early 1987,\cite{ShenZhuPRB1987}
SFG-VS has been widely used to investigate various molecular
interfaces, including vapor/liquid, liquid/liquid, air(vacuum
gas)/solid, and liquid/solid interfaces, because SFG-VS, as one of
the second order nonlinear optical processes, is interface selective
and is sensitive to submonolayer changes at the molecular
interface.\cite{shen1989nature,cdbain:jcsf1995,eisenthal:review,
Shen1999JPCB103p3292,Shultz-internationalreview,buck:review2001,
chenz:review:somorjai,Richmond:cr102:2693,HongfeiIRPCreview} Since
the very beginning, quantitative interpretation of the SFG-VS
spectra has been used to derive molecular orientational structure
and order of the molecular interfaces under
study.\cite{ShenPRL1987,Shen-methanolPRL1991}

The key for such quantitative interpretation lies on the ability to
obtain or estimate the ratios between different microscopic
polarizability $\beta _{i'j'k'}$ tensor elements of a chemical group
in a molecule.\cite{HongfeiIRPCreview,Shen-5CT,RichmondJPCBp16846}
So far, almost all of the efforts have focused on the stretching
vibrational modes. The approaches employed to obtain the $\beta
_{i'j'k'}$ tensor ratios are all based on the following general
relationship between the second order hyperpolarizability tensors
($\beta_{i'j'k'}^{q}$) for a particular vibrational mode $q$ to the
Raman polarizability derivative tensors
($\partial\alpha_{i'j'}/\partial Q_{q}$, usually denote as
$\alpha'_{i'j'}$) and dipole moment derivative tensors
($\partial\mu_{k'}/\partial Q_{q}$, usually denote as $\mu'_{k'}$)
of the $q$th mode, i.e.,\cite{ShenBook,HongfeiIRPCreview,Shen-5CT}

\begin{eqnarray}
\beta_{i'j'k'}^{q} &=&
-\frac{1}{2\epsilon_{0}\omega_{q}}\frac{\partial\alpha_{i'j'}}{\partial
Q_{q}}\frac{\partial\mu_{k'}}{\partial Q_{q}}\label{beta}
\end{eqnarray}

\noindent Here $(i'j'k')$ represents the molecular coordinates
system, $\omega_{q}$ and $Q_{q}$ are the vibrational frequency, and
the normal coordinates of the $q$th vibrational mode of the
molecule, respectively. Therefore, according to Eq.\ref{beta}, if
the proper ratios between different $\alpha'_{i'j'}$ terms and the
ratios between different $\mu'_{k'}$ terms are known, the ratios of
$\beta_{i'j'k'}^{q}$ elements can be readily obtained.

Among the approaches to obtain $\beta_{i'j'k'}^{q}$ tensor ratios,
which shall be compared in detail in Section II, the bond additivity
model is based on symmetry analysis along with the Raman bond
polarizability derivative theory and the bond moment
theory.\cite{HiroseJCP96p997,HiroseJPC97p10064,YRShenPRE62p5160,
HongfeiIRPCreview} In some cases, it has been proven successful in
quantitative interpretation of the symmetric stretching (ss)
vibrational mode in SFG-VS studies.\cite{ShenPRL1987,
Shen-methanolPRL1991,HiroseJCP96p997,HiroseJPC97p10064,DuQuanPRL1993,
YRShenPRE62p5160,AMoritaCP258p371,HongfeiCJCPWater,HongfeiJCPWaterLong,
YRShenJCP114p1837,BaldelliJAP2004,HFWangJPCB108p7297,HFWangJPCB109p14118,
HongfeiIRPCreview} However, it often failed to quantitatively
predict the SFG spectral intensity relationship between the
symmetric stretching (ss) and asymmetric stretching (as) vibrational
modes, as shall be discussed in detail in Section II. Therefore, the
limitations and effectiveness of the bond additivity model still
needs to be investigated.

In this report, we shall employ an complete empirical approach to
the bond additivity model by using experimental IR and Raman spectra
to correct the discrepancies between the $\alpha'_{i'j'}$ and
$\mu'_{k'}$ tensors of the molecular groups. The effectiveness of
this approach shall be demonstrated with the ss and as modes of the
CH$_{3}$ group in CH$_{3}$OH and CH$_{3}$CD$_{2}$OH molecules. We
shall show that with the experimentally corrected $\alpha'_{i'j'}$
and $\mu'_{k'}$ tensor relationships, the $\beta_{i'j'k'}$ tensor
ratios calculated from Eq.\ref{beta} can be used successfully to
quantitatively interpret detailed polarization dependence in SFG-VS
spectra of the vapor/CH$_{3}$OH and vapor/CH$_{3}$CD$_{2}$OH
interfaces. This empirical approach overcomes the limitations of the
current simple bond additivity model, and it also provides a
practical roadmap for its application in SFG-VS studies of molecular
interfaces. This approach also demonstrates how to use molecular IR
and Raman spectra in the bulk liquid or gaseous phases for
quantitative understanding of the SFG-VS vibrational spectra of the
same molecule at the interface, and \textit{vise versa}.

\section{Background}

Here we briefly compare the different approaches for obtaining
$\beta_{i'j'k'}^{q}$ tensor ratios, and discuss how and why the bond
additivity model failed to predict the SFG spectral intensity
relationship between the symmetric stretching (ss) and asymmetric
stretching (as) vibrational modes

\subsection{Three approaches for the $\beta_{i'j'k'}^{q}$ tensor ratios}

As a 3rd rank tensor, $\beta_{i'j'k'}^{q}$ has $3\times3\times3=27$
tensor elements in total. In the most ideal case, the ratios between
all of these tensor elements should be determined \textit{apriori}.
Recently, Hore \textit{et al.} proposed a whole molecule approach
based on a general scheme for \textit{ab initio} calculation of the
vibrational hyperpolarizability of any IR- and Raman-active mode,
regardless of the molecular symmetry or complexity of the structure,
and application of this approach to the -OSO$_{3}$ headgroup of a
surfactant molecule at the air/water interface was also
demonstrated.\cite{RichmondJPCBp16846} With the tremendous
advancement of the capability and availability of the \textit{ab
initio} computation methods, the advantage of this whole molecule
approach is apparent. However, because this approach is purely
computational, its general accuracy can be affected by the
limitations of the current \textit{ab initio} computational methods
in reproducing the Raman and IR spectra intensities, as well as in
dealing with the problem of mode couplings, such as Fermi
resonances. Nevertheless, this attempt does provide an promising
alternative solution to the limitations of other existing
approaches.

Unlike the whole molecule approach mentioned above, these other
approaches all employed the local mode assumption of the particular
molecular group. The bond additivity model even further employed the
local mode assumption of the particular bonds. With the knowledge of
the symmetry of the molecular groups, the number of non-zero
$\beta_{i'j'k'}^{q}$ tensor elements can be significantly reduced;
accordingly, the number of non-zero $\alpha'_{i'j'}$ elements of the
symmetry group can also be significantly
reduced.\cite{ShenBook,BuckinghamJOSAB1998} For example, the
CH$_{3}$ and CH$_{2}$ groups can be treated with $\textit{C}_{3v}$
and $\textit{C}_{2v}$ symmetries, respectively. Thus, their non-zero
$\beta_{i'j'k'}^{q}$ elements for the stretching vibrational modes
are reduced to 11 and 7,
respectively.\cite{ShenBook,BuckinghamJOSAB1998} There have been
discussions on whether the local mode treatment of the CH$_{3}$ and
CH$_{2}$ groups with $\textit{C}_{3v}$ and $\textit{C}_{2v}$
symmetries, respectively, is
valid.\cite{Bain:Lang:1993,CDBainJPCB1998} However, such concerns
have not been a serious issue because it has been generally accepted
in molecular spectroscopy textbook that `...in the case of C-H
stretches, the high frequency of the local vibration of the C-H bond
tends to uncouple that motion from that of the rest of the
molecule'. \cite{MchaleBook} In addition, slight deviation from
$\textit{C}_{3v}$ symmetry of the CH$_{3}$ group in actual molecules
can be treated with small perturbations, as discussed by Hirose
\textit{et al.}\cite{HiroseJCP96p997}.

\begin{center}
\begin{figure}[h]
\includegraphics [width=4.0cm]{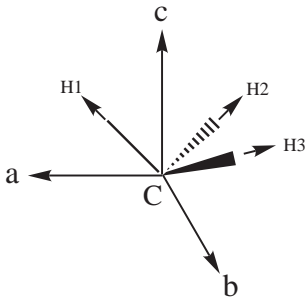}
\caption{Molecule-fixed(abc) axis for $C_{3v}$ symmetry. Illustrated
with the $CH_3$ group.}\label{CH3Coordinates}
\end{figure}
\end{center}

One such approach used experimental Raman depolarization ratio to
determine the $\beta_{i'j'k'}^{q}$ ratios of the symmetric
stretching (ss) mode of $C_{3v}$ and the stretching mode of
$C_{\infty v}$ groups, because both groups have only two independent
$\beta_{i'j'k'}^{q}$ terms for the ss mode under the local group
mode assumption, i.e., there is only one $\beta_{i'j'k'}^{q}$ ratio
$R={\beta_{aac}}/{\beta_{ccc}}={\beta_{bbc}}/{\beta_{ccc}}$ need to
be determined.\cite{DinaJPC1994NA,HongfeiIRPCreview} The coordinates
$(a,b,c)$ is define as in Fig.\ref{CH3Coordinates}. This direct
Raman polarization ratio method worked because according to
Eq.\ref{beta},
$R={\alpha'_{aa}}/{\alpha'_{cc}}={\alpha'_{bb}}/{\alpha'_{cc}}$ for
the ss mode of $C_{3v}$ and the stretching mode of $C_{\infty v}$
groups, and it can be obtained from Raman experimental
depolarization measurement.\cite{HongfeiIRPCreview} This approach
has been widely used and worked successfully because it is purely
empirical, and it solely depends on the reliability of the
particular Raman depolarization
measurement.\cite{DinaJPC1994NA,HongfeiIRPCreview,
JohnsonJPCBpaper,RaoEisenthalJPCB2005,GanWeiSymmetryAnalysis}

One may surmise that even though the group itself may not strictly
observe the $C_{3v}$ or $C_{\infty v}$ symmetry, because this $R$
value is the effective value obtained from the Raman spectral
features with the assumption of $C_{3v}$ or $C_{\infty v}$ symmetry,
it should dutifully reproduce back to the same features of the Raman
spectra with the same assumption of symmetry. Because this $R$ value
is strictly the same for the corresponding $\beta_{i'j'k'}^{q}$
ratio according to Eq. \ref{beta}, it is not hard to see that with
this $R$ value and the same assumption of symmetry, the related
features in the SFG-VS spectral can be reproduced. This is why this
approach has worked so well for the ss mode of the $C_{3v}$ and the
stretching mode of the $C_{\infty v}$ symmetry
groups.\cite{DinaJPC1994NA,HongfeiIRPCreview,
JohnsonJPCBpaper,RaoEisenthalJPCB2005}

However, this approach can not be applied to $C_{2v}$ group, which
has three instead of two independent $\beta_{i'j'k'}^{q}$ terms for
its ss mode.\cite{HongfeiIRPCreview} Furthermore, this approach does
not deal with the ratio between the $\beta_{i'j'k'}^{q}$ of
symmetric and asymmetric stretching modes of the $C_{3v}$ molecular
group. Therefore, it can not be used to address the relative SFG
intensity of the ss and as modes of the $C_{3v}$ molecular group.

An useful alternative method in the SFG-VS literature is the widely
used bond additivity model, which can give $\beta_{i'j'k'}^{q}$
ratios of the $C_{3v}$, $C_{2v}$, $C_{\infty v}$ and other symmetry
groups, including both of their ss and as
modes.\cite{ShenPRL1987,Shen-methanolPRL1991,HiroseJCP96p997,
HiroseJPC97p10064,DuQuanPRL1993,YRShenPRE62p5160,AMoritaCP258p371,
HongfeiCJCPWater,HongfeiJCPWaterLong,YRShenJCP114p1837,BaldelliJAP2004,
HFWangJPCB108p7297,HFWangJPCB109p14118,HongfeiIRPCreview} The bond
additivity model can also be called the bond polarizability
derivative model. As Hore \textit{et al.} pointed out
recently,\cite{RichmondJPCBp16846} the key to the bond additivity
model is that through symmetry analysis the polarizability tensors
of the individual bond stretches ($C_{\infty v}$) are coupled to
produce the normal mode coordinate of a particular molecular group
($C_{3v}$, $C_{2v}$, etc.).\cite{HiroseJCP96p997,HiroseJPC97p10064}
There are two ways to determine the polarizability tensors ratios of
the individual bond stretches, i.e.
$r=\beta_{\xi\xi\zeta}/\beta_{\zeta\zeta\zeta}=\beta_{\eta\eta\zeta}/\beta_{\zeta\zeta\zeta}$,
where $(\xi\eta\zeta)$ is the single bond fixed coordinates with
$\zeta$ as the primary axis of the single bond. One way is through
theoretical (\textit{ab initio}) calculation of the Raman tensors of
the single bond,\cite{GoughJCP1989} and the other way is through
experimentally measured Raman depolarization ratio, or Raman
intensity ratio between the ss and as modes,\cite{HiroseJCP96p997,
HiroseJPC97p10064,HongfeiIRPCreview} on the basis that the single
bond posses $C_{\infty v}$ symmetry and that
$r=\alpha'_{\xi\xi}/\alpha'_{\zeta\zeta}=\alpha'_{\eta\eta}/\alpha'_{\zeta\zeta}$
for a single bond according to Eq.\ref{beta}.

The detailed formula and the effectiveness, as well as some
limitations, of the bond polarizability model was critically
reviewed recently.\cite{HongfeiIRPCreview} So far, this model have
been used for study of H$_{2}$O molecule at the air/water
interfaces,\cite{DuQuanPRL1993,AMoritaCP258p371,HongfeiCJCPWater,
HongfeiJCPWaterLong} -CH$_{2}$- groups\cite{YRShenPRE62p5160,
BaldelliJAP2004,HFWangJPCB108p7297,HFWangJPCB109p14118} and
occasionally -CH$_{3}$ groups\cite{ShenPRL1987,Shen-methanolPRL1991,
HiroseJPC97p10064,YRShenJCP114p1837} at various molecular
interfaces. The unique success of the bond additivity model has been
on H$_{2}$O molecules and -CH$_{2}$- groups to which direct Raman
polarization ratio method can not be applied as mentioned above. In
terms of -CH$_{3}$ or other C$_{3v}$ groups, the bond additivity
model is essentially the same thing as the direct raman
depolarization ratio method for the ss mode. Another unique
advantage of the bond additivity model is that it can give the
$\beta_{i'j'k'}^{q}$ ratio between the ss and as modes for the
C$_{3v}$ group.\cite{HFWangJPCB109p14118,HongfeiIRPCreview} However,
some intrinsic weakness of the bond additivity model has limited its
effectiveness on interpretation of the relative intensity of the ss
and as modes in actual SFG-VS spectra. This is discussed in detail
as in the followings.

\begin{center}
\begin{figure*}[!htb]
\includegraphics[height=9.0cm,width=7.0cm]{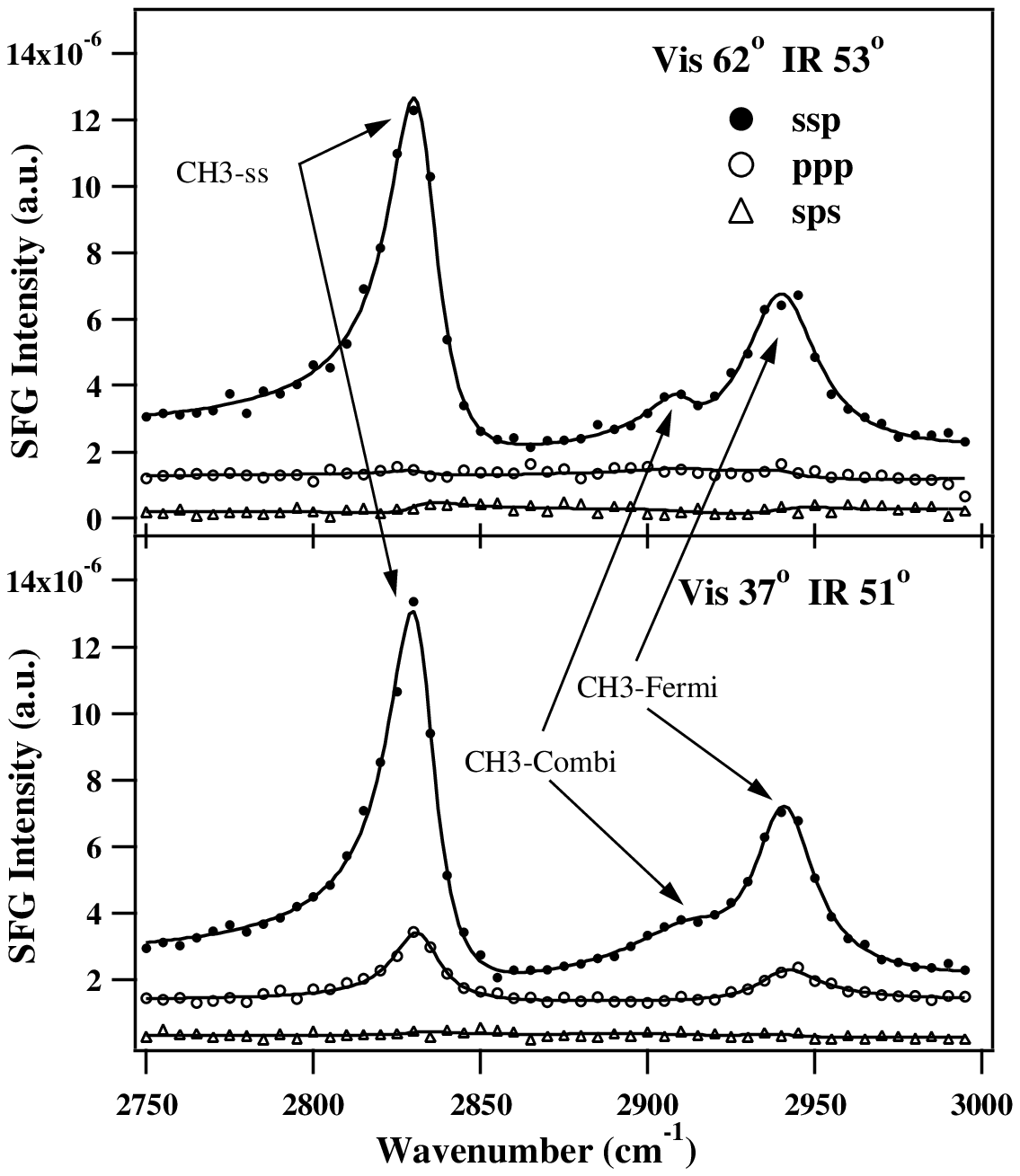}
\hspace{0.5cm}%
\includegraphics[height=9.0cm,width=8.0cm]{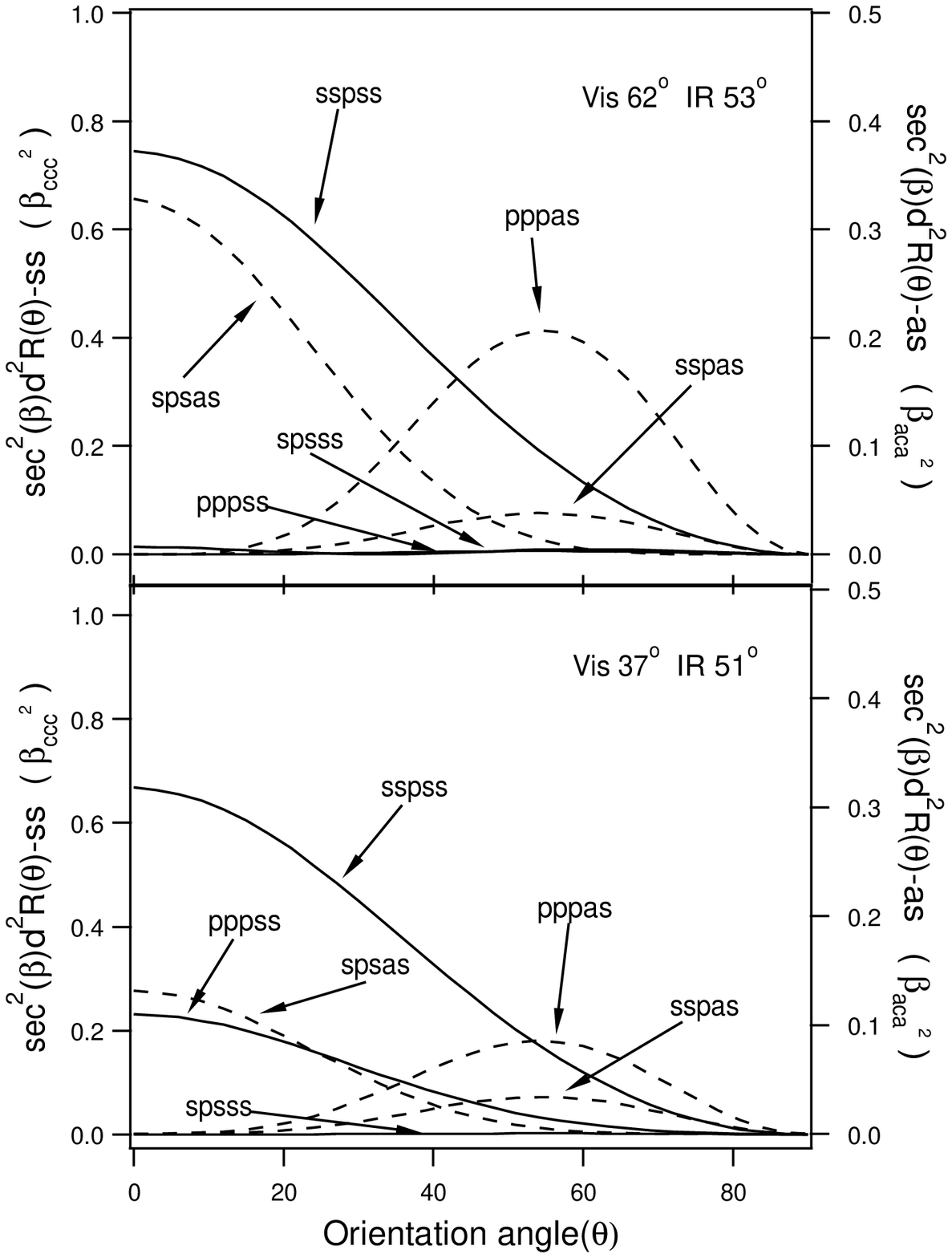}
\caption{SFG spectra of the air/methanol interface in two different
experimental configurations with different incident angles. The
$\beta$ in $\sec^{2}(\beta)$ is the angle between the SFG signal
beam and the surface normal. It is determined by the visible and IR
incident angles. The SFG intensity is proportional to
$\sec^{2}(\beta)d^{2}R(\theta)$ in units of $\beta_{ccc}^{2}$ or
$\beta_{aca}^{2}$ for the ss and as modes, respectively. The solid
lines in the two figure on the left are fittings with Lorentzian
lineshapes.}\label{MethanolSFG}
\end{figure*}
\end{center}

\begin{center}
\begin{figure*}[!htb]
\includegraphics[width=6.0cm]{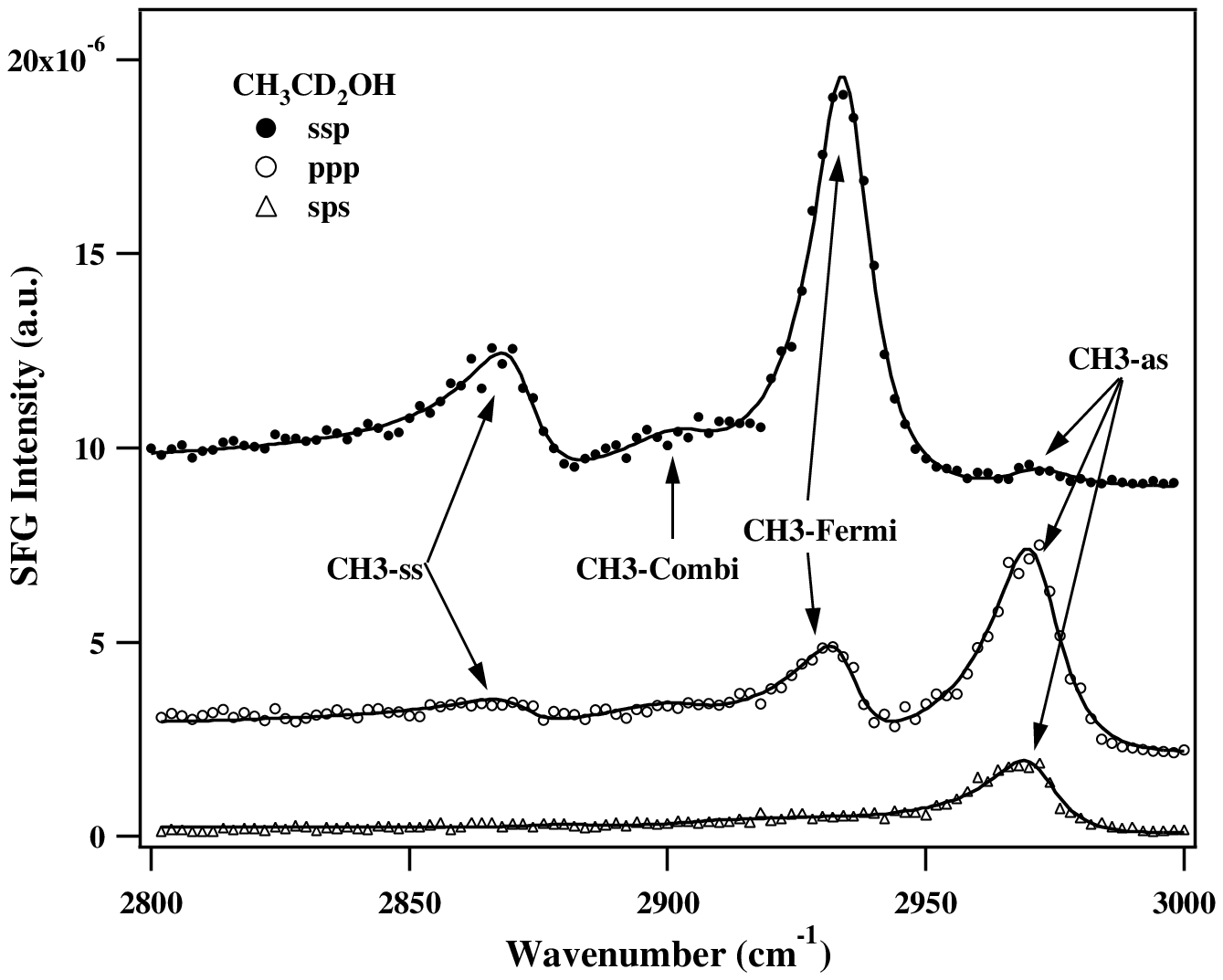}
\hspace{1.5cm}
\includegraphics[width=7.0cm]{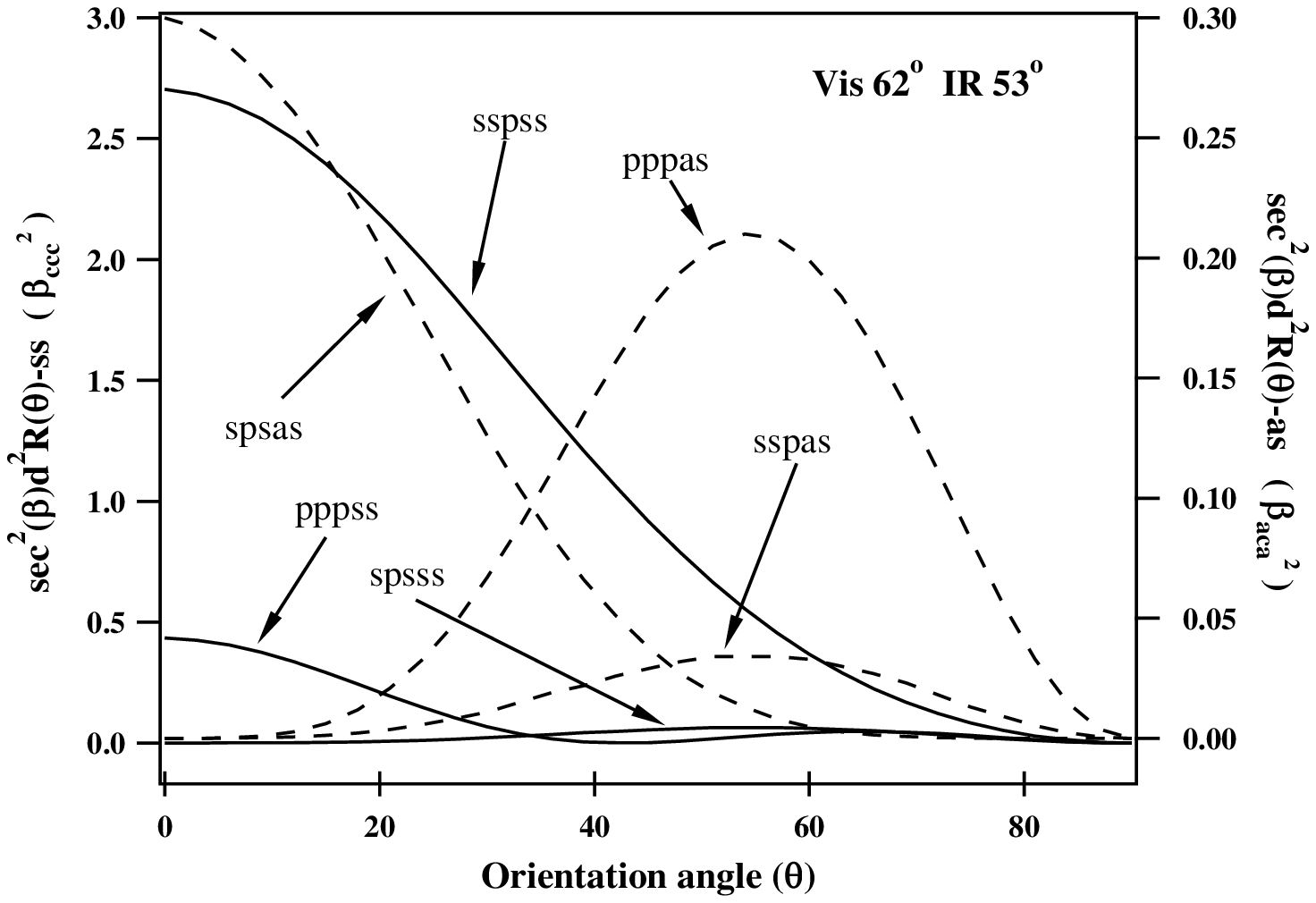}
\caption{SFG-VS spectra of the vapor/CH$_{3}$CD$_{2}$OH interface in
ssp, ppp and sps polarizations with incident angles of
Vis=62$^{\circ}$ and IR=53$^{\circ}$ in the SFG experiment. Because
the CH$_{2}$ group is deuterated, therefore, all spectral features
belong to the CH$_{3}$ group. The solid lines are fittings with
Lorentzian lineshapes.}\label{EthanolSpectra}
\end{figure*}
\end{center}

The bond additivity model uses the Raman bond polarizability
derivative tensors $\alpha'_{i'j'}$ and the bond moment derivative
tensors $\mu'_{k'}$ to calculate $\beta_{i'j'k'}$ tensors according
to Eq.\ref{beta}.\cite{HiroseJPC97p10064,HongfeiIRPCreview} In Raman
and IR spectroscopy studies, the bond polarizability theory and the
bond moment theory have been used to interpret the IR and Raman
spectral intensity.\cite{DixitASR18p373} As we have known, even
though the Raman bond polarizability derivative model have worked
well with the intensities of the ss modes, it has not been very
effective on the relative intensity between the ss and as modes of
the same molecular
group.\cite{YoshinoJMS2p213,ASRcite5,ASRcite34,ASRcite136,DixitASR18p373}
Furthermore, the simple bond moment hypothesis, also called the
zero-order bond moment theory, as used in the bond additivity model
in SFG-VS, has not been successful in interpretation of IR
intensities.\cite{Steele1964QR18p21,DixitASR18p373,HisatsuneJCP23p487,
EggersJPC59p1124,HornigJPC59p1133} This is because the simple bond
moment theory essentially neglects all coupling effect between the
single bonds even within the same molecular group. Therefore,
modified bond moment theory which includes such and other coupling
effects has been proposed,\cite{DixitASR18p373} but such modified
theory becomes complicated by introduced many unknown parameters in
order to address the coupling terms. In short, these facts certainly
limit the effectiveness of the bond additivity model for
quantitative interpretation of the SFG-VS spectra.

However, with all these problems, two of the important and
successful features of the bond additivity model should not be
overlooked. Firstly, the bond additivity model rooted deeply into
the concept of molecular symmetry analysis, which is the basis for
the polarization selection rules and has provided clear physical
picture for understanding of the polarization and orientational
dependence in
SFG-VS,\cite{HFWangJPCB108p7297,HFWangJPCB109p14118,HongfeiIRPCreview}
as well as all other spectroscopic
techniques.\cite{SymmetrySpectroscopyBook} Secondly, the direct
Raman polarization ratio method has been quantitatively successful
in many of the SFG-VS
applications.\cite{DinaJPC1994NA,HongfeiIRPCreview,
JohnsonJPCBpaper,RaoEisenthalJPCB2005,GanWeiSymmetryAnalysis} Since
a full computational solution for quantitative interpretation of the
SFG-VS spectra is yet not within our reach as mentioned above, it
can certainly be beneficial to extend the bond additivity model if
we can use the core concept for its successes, and try to directly
address the limitations in the consisting Raman and IR theories.

\subsection{Failure of the bond additivity model}

The failure of the bond additivity model came to our attention in
our recent study of SFG-VS spectra of vapor/alcohol interfaces of
C$_1$-C$_8$ alcohols in the CH stretching vibration region between
2800 to 3000cm$^{-1}$.\cite{HFWangJPCB109p14118,HongfeiIRPCreview}
The fact has been long known that there has been no observation of
the $CH_3$-as peak, which should be around 2970cm$^{-1}$, in the
SFG-VS spectra of the vapor/methanol
interface;\cite{Shen-methanolPRL1991,
Laubereau1993Methanol,HuangJYPRE1994,ShenCPL1995Alcohol,
AllenJPCB2003Methanol,Lurong1,ChenHua3,KimJPCB2005,HFWangJPCB109p14118}
while for longer chain alcohols ($C_2-C_8$ alcohols), the $CH_3$-as
peak is apparently very strong in their SFG-VS
spectra.\cite{ShenCPL1995Alcohol,HFWangJPCB109p14118,KimJPCB2005}
Even though the bond additivity model can seemingly explain why the
longer chain alcohols can have stronger $CH_3$-as peak in the SFG-VS
spectra of their vapor/liquid interfaces than that of the
vapor/methanol interface,\cite{HongfeiIRPCreview} it also predict
that the $CH_3$-as mode of methanol should be observable in the sps
SFG-VS spectrum.

Fig.\ref{MethanolSFG} presents the SFG-VS spectra and the calculated
polarization dependence against the CH$_{3}$ orientational angle
$\theta$ from the interface normal using the bond additivity model
in the ssp, ppp, and sps polarizations for two different
experimental configurations. The SFG-VS intensity is proportional to
$\sec^{2}(\beta)d^{2}R(\theta)$, and the detail of calculation has
been presented
previously.\cite{HongfeiIRPCreview,HFWangJPCB109p14118} It is to be
noted that calculation of the ss mode intensity uses
$\beta_{ccc}^{2}$ as unit, and the as mode uses $\beta_{aca}^{2}$ as
unit. It is clear that for both experimental configurations,
$CH_3$-as mode around 2970cm$^{-1}$ can not be observed in the
SFG-VS spectra. This tells us that the missing of the $CH_3$-as mode
peak is not something accidental. With the bond additivity model,
methanol has $\beta_{ccc}/\beta_{aca}=1.1$ for its CH$_{3}$ group,
as calculated in Section III. Therefore, from the calculation
results in Fig.\ref{MethanolSFG}, $CH_3$-as should be experimentally
observable in at least one of the three polarizations in comparison
to the intensities of the $CH_3$-ss mode. Recent measurements have
determined that for vapor/methanol interface the $CH_3$ orientation
is close to the interface
normal.\cite{Lurong1,ChenHua3,GanweiCPLNull,HongfeiIRPCreview} With
this CH$_{3}$ group orientation, the bond additivity model predicted
that the $CH_3$-ss mode in the ppp polarization can be observed in
the second experimental configuration (VIS=37$^{\circ}$ and
IR=51$^{\circ}$), and this spectral feature was indeed observed with
the expected intensity as indicated in Fig.\ref{MethanolSFG}. This
fact certainly suggests the effectiveness of the bond additivity
model calculations for the $CH_3$-ss mode of methanol molecule.
Accordingly, if the bond additivity calculation is also effective
for the as mode, the $CH_3$-as mode intensity in the sps
polarization should be more than 1/3 of that of the $CH_3$-ss mode
in the ssp polarization of the first experimental configuration.
However, no such $CH_3$-as mode has been ever
observed.\cite{Shen-methanolPRL1991,
Laubereau1993Methanol,HuangJYPRE1994,ShenCPL1995Alcohol,
AllenJPCB2003Methanol,Lurong1,ChenHua3,KimJPCB2005,HFWangJPCB109p14118}
This clearly demonstrates the failure of the bond additivity model
in dealing with the $CH_3$-as mode, even though the intensities of
the $CH_3$-ss and $CH_3$-ss-Fermi modes can be quantitatively
explained as in Fig.\ref{MethanolSFG}.

In comparison, Fig.\ref{EthanolSpectra} presents the SFG-VS spectra
and the bond additivity model calculation of
vapor/CH$_{3}$CD$_{2}$OH interface with the incident angles
Vis=62$^{\circ}$ and IR=53$^{\circ}$ in the SFG-VS experiment. It is
clear that the $CH_3$-as mode around 2970cm$^{-1}$ in the ppp
polarization is even much bigger than the $CH_3$-ss mode around
2870cm$^{-1}$ in the ssp polarization. This is consistent with the
bond additivity model value of $\beta_{ccc}/\beta_{aca}=0.30$ for
the CH$_{3}$ group of CH$_{3}$CD$_{2}$OH molecule, as calculated in
Section IV. Then, why the bond additivity model works for the
$CH_3$-as mode of CH$_{3}$CD$_{2}$OH but not that of CH$_{3}$OH?

Even though the bond additivity model and the direct Raman
polarization ratio method have been apparent success in quantitative
interpretation of the polarization dependence and orientational
analysis of the CH$_{3}$ groups from the SFG-VS spectra of its
$CH_3$-ss mode,\cite{Lurong1,ChenHua3,HongfeiIRPCreview} their
inability and discrepancy to address the as mode intensities of the
CH$_{3}$ group in CH$_{3}$CD$_{2}$OH and CH$_{3}$OH molecules are
certainly unpleasant, and put uncertainties to these models.
However, as discussed above, the bond additivity model is still the
most useful model in obtaining the $\beta_{i'j'k'}^{q}$ tensor
ratios. We realized that the effectiveness of these models on the ss
mode of the $C_{3v}$ and the stretching mode of the $C_{\infty v}$
symmetry lies on the fact that their $R$ values are empirically
obtained. On the other hand, the failures of these models also lie
on the fact that those other ratios were not set on the same
empirical basis. In order to address these problems and to have
better understanding of the bond additivity model, we propose an
complete empirical approach in order to correct the failures of the
bond additivity model. This approach includes the full empirical
treatment of the ss vs. as ratios of the raman and IR tensors, in
addition to the successful empirical treatment of the ss mode.

\section{Bond additivity model and its empirical correction}

In a recently review on quantitative treatment on SFG-VS from
molecular interfaces, we discussed related issues on the bond
additivity model in detail.\cite{HongfeiIRPCreview} We pointed out
that the original expression given by Hirose \textit{et
al.}\cite{HiroseJCP96p997, HiroseJPC97p10064} needed to be corrected
for some minor errors, and the correct expressions for the C$_{3v}$,
C$_{2v}$ and C$_{\infty v}$ groups were
presented.\cite{HongfeiIRPCreview} We also presented some good
examples to demonstrate the validity of the bond additivity model
for quantitative analysis of SFG-VS spectra, especially for
interpretation of the ss modes of the C$_{3v}$, C$_{2v}$ and
C$_{\infty v}$ groups, and we also pointed out that further detailed
examinations of this model were still
needed.\cite{HongfeiIRPCreview}

Therefore, here we shall try to keep the successful part of the bond
additivity model, which is to use the experimental Raman
depolarization ratio of the ss mode to obtain the single bond Raman
polarizability derivative tensor ratio $r$, and try to derive the
expressions for the Raman and IR intensity expressions for the ss
and as mode of the C$_{3v}$ group, in order to obtain the
corrections factors for the $\alpha'_{i'j'}$ ratios and $\mu'_{k'}$
ratios between the ss and as modes from the experimental Raman and
IR spectra, respectively. We hope this approach can produce the
corrected $\beta_{i'j'k'}^{q}$ tensor ratios between the ss and as
modes, and can be used to quantitatively address the above discussed
problems in the SFG-VS spectra of the vapor/CH$_{3}$OH and
vapor/CH$_{3}$CD$_{2}$OH interfaces.

For the stretching vibrational modes of a molecular group with
C$_{3v}$ symmetry, there are a single symmetric mode ($A_{1}$) and a
doubly degenerated asymmetric mode ($E$), with the following
relationships between the $\beta_{i'j'k'}$, $\alpha'_{i'j'}$ and
$\mu'_{k'}$ tensors when there is no electronic
resonances.\cite{NafieJCP1972,Longbook,HongfeiIRPCreview}

For the $A_{1}$ (symmetric) mode:
\begin{eqnarray}
\mbox{SFG-VS \ \ } &&
\beta_{aac}=\beta_{bbc}=R\beta_{ccc}\label{SFGss}\\
\mbox{Raman: \ \ } &&
\alpha'_{aa}=\alpha'_{bb}=R\alpha'_{cc}\label{Ramanss} \\
\mbox{IR: \ \ } && \mu'_{c}
\end{eqnarray}

For the E (asymmetric) mode:
\begin{eqnarray}
\mbox{SFG-VS \ \ } &&
\beta_{aca}=\beta_{bcb}=\beta_{caa}=\beta_{cbb}\nonumber\\
&&\beta_{aaa}=-\beta_{bba}=-\beta_{abb}=-\beta_{bab}\label{SFGas}\\
\mbox{Raman: \ \ } && \alpha'_{ac}=\alpha'_{ca}=\alpha'_{bc}=\alpha'_{cb}\nonumber\\
&& \alpha'_{aa}=-\alpha'_{bb}=-\alpha'_{ab}=-\alpha'_{ba}\label{Ramanas}\\
\mbox{IR: \ \ } && \mu'_{a}=\mu'_{b}
\end{eqnarray}

It is to be noted that the $\alpha'_{aa}$ and $\alpha'_{bb}$ tensors
of the symmetric mode are different from the $\alpha'_{aa}$ and
$\alpha'_{bb}$ tensors of the asymmetric mode.

For a rotationally isotropic interface, the four asymmetric
$\beta_{i'j'k'}$ terms, namely
$\beta_{aaa}=-\beta_{bba}=-\beta_{abb}=-\beta_{bab}$, are dumb
tensors, which would not appear in the non-zero elements of the 3rd
rank macroscopic susceptibility $\chi_{ijk}$
tensors.\cite{HongfeiIRPCreview} Therefore, if
$R=\beta_{aac}/\beta_{ccc}$ is known from the direct Raman
depolarization ratio method, in addition, only the ratio
$\beta_{ccc}/\beta_{aca}$ need to be obtained for all useful
$\beta_{i'j'k'}$ tensor
ratios.\cite{HiroseJPC97p10064,HongfeiIRPCreview} According to
Eq.\ref{beta}, one has,

\begin{eqnarray}
\frac{\beta_{ccc}}{\beta_{aca}}=\frac{\omega_{asym}}{\omega_{sym}}
\frac{\alpha'_{cc}}{\alpha'_{ac}}\frac{\mu'_{c}}{\mu'_{a}}\label{BetaRatio}
\end{eqnarray}

Therefore, with the frequencies of the symmetric and asymmetric
modes ($\omega_{sym}$ and $\omega_{asym}$, respectively) known, only
the Raman tensor ratio $\alpha'_{cc}/\alpha'_{ac}$ and the IR tensor
ratio $\mu'_{c}/\mu'_{a}$ need to be known to make the calculation
of $\beta_{ccc}/\beta_{aca}$.

The IR tensor ratio $\mu'_{c}/\mu'_{a}$ can be directly obtained
from the IR intensity ratio of the symmetric and asymmetric modes.
However, in the simple bond additivity model, the
$\mu'_{c}/\mu'_{a}$ ratio was not experimentally
corrected.\cite{HiroseJPC97p10064,HongfeiIRPCreview}

\begin{eqnarray}
\frac{I_{sym}^{IR}}{I_{asym}^{IR}}=\left(\frac{\mu'_{c}}{\mu'_{a}+\mu'_{b}}\right)^2
=\left(\frac{\mu'_{c}}{2\mu'_{a}}\right)^2\label{IR}
\end{eqnarray}

In order to get $R$ and $\alpha'_{cc}/\alpha'_{ac}$, the following
Raman polarizability theory need to be invoked.

When vertically linear polarized light is used for the Raman
experiment, the following relationships for the Raman depolarization
ratio and Raman intensity ratio in the parallel polarization to the
individual Raman polarizability derivative tensors
($\alpha'_{i'j'}$) are generally valid for all molecular
groups.\cite{Long1977,Longbook,GriffithsJCP1972}

\begin{eqnarray}
\rho&=&\frac{3\gamma^2}{45\alpha^2+4\gamma^2}\label{RamanDPRatio}\\
\frac{I_{sym}^{Raman}}{I_{asym}^{Raman}}&=&\frac{\alpha_{sym}^2+
\frac{4}{45}\gamma_{sym}^2}{\alpha_{asym}^2+\frac{4}{45}\gamma_{asym}^2}\label{RamanRatio}
\end{eqnarray}

\noindent with the definition of

\begin{eqnarray}
\alpha&=&\frac{1}{3}(\alpha'_{aa}+\alpha'_{bb}+\alpha'_{cc})\label{alpha}\\
\gamma^2&=&\frac{1}{2}[{(\alpha'_{aa}-\alpha'_{bb})}^2+{(\alpha'_{bb}-\alpha'_{cc})}^2\nonumber \\
&&+{(\alpha'_{cc}-\alpha'_{aa})}^2+6({\alpha'_{ab}}^2+{\alpha'_{bc}}^2+{\alpha'_{ca}}^2)]\label{gamma}
\end{eqnarray}

\noindent The molecular coordinates are defined as in
Fig.\ref{CH3Coordinates}. If natural light is used for the Raman
spectra measurement, the factor $4/45$ used in Eq.\ref{RamanRatio}
should be $7/45$, and the depolarization ratio becomes
$\rho=6\gamma^{2}/(45\alpha^{2}+7\gamma^{2})$.\cite{Long1977,Longbook,GriffithsJCP1972}

In this case, if one puts Eq.\ref{Ramanss} into
Eq.\ref{RamanDPRatio}, because there are only three non-zero
symmetric tensors, a unique $R$ value (generally $1\leq R \leq 4$
for C$_{3v}$ groups, and $0\leq R \leq 1$ for C$_{\infty v}$ groups)
can be obtained from the symmetric tensors through the experimental
value of the Raman depolarization ratio $\rho$ of the $C_{3v}$
symmetric mode.\cite{Shen-5CT,HongfeiIRPCreview}

\begin{eqnarray}
\rho_{sym}&=&\frac{3\gamma^{2}_{sym}}{45\alpha^{2}_{sym}+4\gamma^{2}_{sym}}\nonumber\\
&=&\frac{3}{4+5[(1+2R)/(R-1)]^2}\label{rhodefinition}
\end{eqnarray}

Here the Eq.\ref{rhodefinition}  for obtaining $R$ is the so called
direct Raman depolarization ratio method because $R$ directly
satisfies Eq.\ref{SFGss} according to
Eq.\ref{beta}.\cite{DinaJPC1994NA,HongfeiIRPCreview}

However, since the raman depolarization ratio for the asymmetric
mode $\rho_{asym}=3/4$ is always a constant, no knowledge of the
asymmetric tensors can be learned from the direct Raman
depolarization ratio method. Therefore, one now have to employ the
bond additivity model to make such connection through
Eq.\ref{RamanRatio}.

In the bond additivity model, the single bond property is used to
calculate the property of the whole group. Following Hirose
\textit{et al.}'s formulation,\cite{HiroseJPC97p10064,
WubaohuaThesis} the Raman polarizability derivative of the single
bond can be related to the following three non-zero tensor elements
with the following relationship,

\begin{eqnarray}
\frac{\partial\alpha_{\xi\xi}}{\partial
\Delta\zeta}=\frac{\partial\alpha_{\eta\eta}}{\partial\Delta\zeta}
=r\frac{\partial\alpha_{\zeta\zeta}}{\partial\Delta\zeta}\label{BondAlpha}
\end{eqnarray}

\noindent in which $\zeta$ is the primary axis of the single bond
coordinates, and the stretching vibration displacement $\Delta\zeta$
from its equilibrium position is along the single bond. Generally,
for the single bond, $0\leq r\leq 1$.\cite{HongfeiIRPCreview} This
is to say, the single bond is treated with $C_{\infty v}$ symmetry.
In the bond additivity
model,\cite{HiroseJCP96p997,HiroseJPC97p10064,
WubaohuaThesis,HongfeiIRPCreview} we define the following,

\begin{eqnarray}
&&a_{0}=(\frac{\partial\alpha_{\zeta\zeta}}{\partial\Delta\zeta})_{0}\nonumber\\
&&m_{0}=(\frac{\partial\mu_{\zeta}}{\partial\Delta\zeta})_{0}\nonumber\\
&&G_{A_{1}}=\frac{1+2\cos{\tau}}{M_A}+\frac{1}{M_B}\nonumber\\
&&G_{E}=\frac{1-\cos{\tau}}{M_A}+\frac{1}{M_B}\nonumber
\end{eqnarray}

\noindent in which $\tau$ is the tetrahedral angle between two
single $A$-$B$ bonds, $M_{A}$ and $M_{B}$ are the atomic masses of
the atom $A$ and $B$ of the $AB_{3}$ with C$_{3v}$ symmetry.
Therefore, $G_{A_{1}}$ and $G_{E}$ are the inverse reduced masses of
the $A_{1}$ and $E$ normal modes, respectively, with the normal mode
coordinates defined as,\cite{HiroseJCP96p997,HiroseJPC97p10064,
WubaohuaThesis}

\begin{eqnarray}
&&Q_{A_{1}}=(\Delta r_{1}+\Delta r_{2}+\Delta r_{3})/(3G_{A_{1}})^{1/2}\nonumber\\
&&Q_{E,a}=(2\Delta r_{1}-\Delta r_{2}-\Delta r_{3})/(6G_{E})^{1/2}\nonumber\\
&&Q_{E,b}=(\Delta r_{2}-\Delta r_{3})/(2G_{E})^{1/2}\nonumber
\end{eqnarray}

\noindent in which $\Delta r_{i}$ is the bond displacement vector
along the direction of the $i$th bond, with the C$_{3v}$ coordinates
defined as in Fig.\ref{CH3Coordinates}. Then according to the bond
additivity model, we have the following expressions for the
$\alpha'_{i'j'}$ and $\mu'_{k'}$ tensors of the C$_{3v}$ group.

For the $A_{1}$ (symmetric) mode:
\begin{eqnarray}
&&\alpha'_{aa}=\alpha'_{bb}\nonumber\\
&&\qquad=\frac{1}{2}a_{0}[(1+r)-(1-r)\cos^2{\tau}]{(3G_{A_{1}})}^\frac{1}{2}\nonumber\\
&&\alpha'_{cc}=a_{0}(\cos^2{\tau}+r\sin^2{\tau}){(3G_{A_{1}})}^\frac{1}{2}\nonumber\\
&&\mu'_{c}=-m_{0}\cos{\tau}{(3G_{A_{1}})}^\frac{1}{2}\nonumber
\end{eqnarray}

For the $E$ (asymmetric) mode:
\begin{eqnarray}
&&\alpha'_{ac}=\alpha'_{ca}=\alpha'_{bc}\nonumber\\
&&\qquad=\alpha'_{cb}=\frac{1}{2}a_{0}(r-1)\sin{\tau}\cos{\tau}{(6G_{E})}^\frac{1}{2}\nonumber\\
&&\alpha'_{aa}=-\alpha'_{bb}=-\alpha'_{ab}\nonumber\\
&&\qquad=-\alpha'_{ba}=\frac{1}{4}a_{0}(1-r)\sin^2{\tau}{(6G_{E})}^\frac{1}{2}\nonumber\\
&&\mu'_{a}=\mu'_{b}=\frac{1}{2}m_{0}\sin{\tau}{(6G_{E})}^\frac{1}{2}\nonumber
\end{eqnarray}

It is to be noted that the above expressions is derived with the
assumption that the $C_{3v}$ group closely assumes a normal
tetrahedral structure. For most the CH$_{3}$ groups under our
investigation, this is generally a good assumption. In order to
simplify calculations, here we use $\cos\tau=-1/3$, which is exact
for the normal tetrahedral angle $\tau\approx109.5^{\circ}$.
Therefore, with $M_{C}=12$ and $M_{H}=1$, we have
$G_{A_{1}}/G_{E}=37/40$.

In order to make the bond additivity model work, the single bond
value of $r$ need to be determined. The $r$ value can be determined
from computations,\cite{GoughJCP1989} or from Raman experimental
measurement.\cite{HiroseJCP96p997,HiroseJPC97p10064,
WubaohuaThesis,HongfeiIRPCreview}. With above $\alpha'_{i'j'}$
expressions put into Eq.\ref{RamanDPRatio} and Eq.\ref{RamanRatio},
we have the following two expressions which can each produce an $r$
value from the experimental Raman depolarization ratio or the Raman
intensity ratio between the ss and the as modes, respectively. For
the Raman measurement with a linearly polarized light, we have,

\begin{eqnarray}
&&\rho_{sym}=\frac{0.75}{1+11.25[(1+2r)/(1-r)]^{2}}\label{rho-r}\\
&&\frac{I_{sym}^{Raman}}{I_{asym}^{Raman}}=\frac{45(1+2r)^2+4(1-r)^2}{32(1-r)^2}\frac{G_{A_{1}}}{G_{E}}\label{Ratio-r}\\
&&\qquad=\frac{45(1+2r)^2+4(1-r)^2}{8(1+8r)^2}\left(\frac{\alpha'_{cc}}{\alpha'_{ac}}\right)^2\label{RatioCorrection}
\end{eqnarray}

\noindent with the following relationships,

\begin{eqnarray}
&&\frac{\alpha'_{cc}}{\alpha'_{ac}}=\frac{1+8r}{2-2r}\left(\frac{G_{A_{1}}}{G_{E}}\right)^\frac{1}{2}\label{BAM1}\\
&&\frac{\mu'_{c}}{\mu'_{a}}=\frac{1}{2}\left(\frac{G_{A_{1}}}{G_{E}}\right)^\frac{1}{2}\label{BAM2}\\
&&\frac{\beta_{ccc}}{\beta_{aca}}=\frac{\omega_{asym}}{\omega_{sym}}\frac{1+8r}{4(1-r)}\frac{G_{A_{1}}}{G_{E}}\label{BAM3}
\end{eqnarray}

Some of the Raman spectra were measured with naturally polarized
light, then the Eq.\ref{Ratio-r} and Eq.\ref{RatioCorrection}
becomes,

\begin{eqnarray}
&&\frac{I_{sym}^{Raman}}{I_{asym}^{Raman}}=\frac{45(1+2r)^2+7(1-r)^2}{56(1-r)^2}\frac{G_{A_{1}}}{G_{E}}\label{Ratio-rNat}\\
&&\qquad=\frac{45(1+2r)^2+7(1-r)^2}{14(1+8r)^2}\left(\frac{\alpha'_{cc}}{\alpha'_{ac}}\right)^2\label{RatioCorrectionNat}
\end{eqnarray}

One would naturally expect that Eq.\ref{rho-r} and Eq.\ref{Ratio-r}
(or Eq.\ref{Ratio-rNat}) would end up with the same $r$ value from
Raman experiment measurement on the same
molecule.\cite{HiroseJPC97p10064} However, experimental data have
shown that it has generally not been the case. Therefore, in the
bond additivity model, Eq.\ref{rho-r} and Eq.\ref{Ratio-r} can not
be used at the same time. This fact has not been clearly recognized
since the first derivation of the bond additivity model by Hirose et
al.\cite{HiroseJPC97p10064} Now the question is whether one of them
can be used, or both are not good at all. As we have known that the
direct Raman polarization ratio method using Eq.\ref{rhodefinition}
involves only with the symmetric Raman tensors of the group, and it
has been widely and successfully
employed.\cite{DinaJPC1994NA,HongfeiIRPCreview,JohnsonJPCBpaper,
RaoEisenthalJPCB2005,GanWeiSymmetryAnalysis} As has been shown
before, the bond additivity model with the single bond $r$ value
from Eq.\ref{rho-r} is equivalent to direct raman depolarization
ratio method with Eq.\ref{rhodefinition}.\cite{HongfeiIRPCreview}
Therefore, it is reasonable to stick with the $r$ value obtained
from Eq.\ref{rho-r}, and then try to experimentally correct the
$\alpha'_{cc}/\alpha'_{ac}$ ratio with Eq.\ref{RatioCorrection} by
using the $r$ value obtained from Eq.\ref{rho-r}. We surmise such
correction directly from the Raman depolarization ratio measurement
and the Raman intensity measurement can correct the value for the
asymmetric tensors, and in the meantime keep the relationship
between the symmetric tensors intact.

Here Eq.\ref{BAM2} from the bond additivity model can be directly
tested with IR spectral measurement using Eq.\ref{IR}. As we have
know from the discussion in Section II, it would generally fail
because this simple bond moment model has not included the coupling
between the single bonds. Therefore, in order to get the correct
$\beta_{ccc}/\beta_{aca}$ ratio according to Eq.\ref{BetaRatio}, the
$\mu'_{c}/\mu'_{a}$ ratio also needs to be corrected with the
measured IR spectra using Eq.\ref{IR}.

We hope by putting the corrected $\alpha'_{cc}/\alpha'_{ac}$ and
$\mu'_{c}/\mu'_{a}$ ratios back into Eq.\ref{BetaRatio}, the
corrected $\beta_{ccc}/\beta_{aca}$ ratio thus obtained can be used
to address the failures of the bond additivity model discussed in
Section IIB. If we can do so, it shall put the uncertain parameters
used in the bond additivity model on a solid empirical basis. In the
next section we shall used the well established IR and Raman data of
liquid CH$_{3}$OH and CH$_{3}$CD$_{2}$OH molecules in the previous
literatures to demonstrate how this empirical approach to the bond
additivity model works. The terms and expressions for the $C_{2v}$
groups are listed in the Appendix.

\section{Results and Discussion}

\subsection{Bond Additivity Model Results}

The Raman depolarization ratio for the ss mode of the CH$_{3}$ group
in CH$_{3}$OH and CH$_{3}$CH$_{2}$OH molecules in bulk liquid at the
room temperature measured with polarized laser light are
$\rho=0.014$ and $\rho=0.053$, respectively.\cite{GriffithsJCP1972}
Photoacoustic stimulated Raman measurement of CH$_{3}$CH$_{2}$OH,
CH$_{3}$CD$_{2}$OH and CD$_{3}$CH$_{2}$OH in the gas phase gave
$\rho=0.060\pm 0.015$ for all three molecules.\cite{LiuSLData} These
values are in good agreement with each other. Because the previous
Raman experiment claimed to have higher accuracy (error of $\rho$
was claimed to be 0.0007), and was measured with bulk liquids, here
we adopt the value $\rho=0.053$ for CH$_{3}$CD$_{2}$OH in this work.
From Eq.\ref{rhodefinition} and Eq.\ref{rho-r}, the $R$ and $r$
values for the CH$_{3}$ group in CH$_{3}$OH and CH$_{3}$CH$_{2}$OH
molecules can be directly calculated.

Therefore, for CH$_{3}$OH, $R=1.7$ and $r=0.28$; while for
CH$_{3}$CH$_{2}$OH, $R=3.4$, and $r=0.026$.

Using the two $r$ values, the following values can be obtained from
the uncorrected bond additivity model for the CH$_{3}$ group in
CH$_{3}$OH and CH$_{3}$CH$_{2}$OH, using the three equations
Eq.\ref{BAM1}-Eq.\ref{BAM3}.

For CH$_{3}$OH, with $r=0.28$, $\omega_{sym}=2836cm^{-1}$ and
$\omega_{asym}=2987cm^{-1}$, we have,

\begin{eqnarray}
\frac{\alpha'_{cc}}{\alpha'_{ac}}=2.2\mbox{ ;} \qquad
\frac{\mu'_c}{\mu'_a}=0.48\mbox{ ;} \qquad
\frac{\beta_{ccc}}{\beta_{aca}}=1.1
\end{eqnarray}

For CH$_{3}$CD$_{2}$OH, with $r=0.026$, $\omega_{sym}=2870cm^{-1}$
and $\omega_{asym}=2976cm^{-1}$, we have,

\begin{eqnarray}
\frac{\alpha'_{cc}}{\alpha'_{ac}}=0.60\mbox{ ;} \qquad
\frac{\mu'_c}{\mu'_a}=0.48\mbox{ ;} \qquad
\frac{\beta_{ccc}}{\beta_{aca}}=0.30
\end{eqnarray}

These values were the same as calculated in the literature, and were
used to make the calculations for Fig.\ref{MethanolSFG} and
Fig.\ref{EthanolSpectra}.\cite{HongfeiIRPCreview}

\subsection{Empirical Corrections of Bond Additivity Model Results}

In order to make empirical corrections of the
$\alpha'_{cc}/\alpha'_{aa}$ and $\mu'_{c}/\mu'_{a}$ ratios in the
bond additivity model, the relative intensities of the CH$_{3}$-ss
and CH$_{3}$-as peaks in the Raman and IR spectra need to be known.

\begin{center}
\begin{table}
\caption{Integrated Raman\cite{GriffithsJCP1972} and
IR\cite{BertieJMS413p333} band intensities of liquid methanol at
298K. All values are normalized to the strongest peak in the Raman
and IR spectra, respectively.\label{Methanoltable}}
\begin{tabular}{llccc}
\hline  &Band &$\nu(cm^{-1})$&FWHH$(cm^{-1})$ & Intensity\\
\hline
Raman & CH$_{3}$-ss &2836 & 20 & 1.00 \\
& CH$_{3}$-Combi &2885 & 39 & 0.16 \\
& CH$_{3}$-Combi &2918 & 28 & 0.26 \\
& CH$_{3}$-Fermi &2944 & 24 & 0.84 \\
& CH$_{3}$-as &2987 & 40 & 0.20 \\
\hline
IR & CH$_{3}$-ss &2833 & 24 & 0.60 \\
& CH$_{3}$-Combi &2872 & 39 & 0.18 \\
& CH$_{3}$-Combi &2911 & 26 & 0.98 \\
& CH$_{3}$-Fermi &2946 & 37 & 1.00 \\
& CH$_{3}$-as &2983 & 47 & 0.70 \\
\hline
\end{tabular}
\end{table}
\end{center}

A detailed band fitting of the Raman spectra from a polarized Raman
measurement of liquid CH$_{3}$OH at 298$K$ were reported by
Griffiths \textit{et al.} in 1972.\cite{GriffithsJCP1972} The
integrated band intensity of the four C-H stretching bands thus
obtained are listed in Table \ref{Methanoltable}. The intensity
ratio between the CH$_{3}$-ss and CH$_{3}$-as is thus
1.00/0.20=5.00, which is used in Eq.\ref{RatioCorrection} to
calculate the corresponding $\alpha'_{cc}/\alpha'_{ac}$.

A detailed band fitting of the IR spectra of liquid CH$_{3}$OH at
298$K$ were reported by Bertie \textit{et al.} in
1997.\cite{BertieJMS413p333} The integrated band intensity of the
C-H stretching bands are also listed in Table \ref{Methanoltable}.
It is to be noted that the Raman and IR band positions and band
widths (Table \ref{Methanoltable}) are in very good agreement with
each other. The intensity ratio between the CH$_{3}$-ss and
CH$_{3}$-as is thus 0.60/0.70=0.86, which is used in Eq.\ref{IR} to
calculate the corresponding $\mu'_{c}/\mu'_{a}$ value.

\begin{center}
\begin{figure}[tb]
\includegraphics [width=7.0cm]{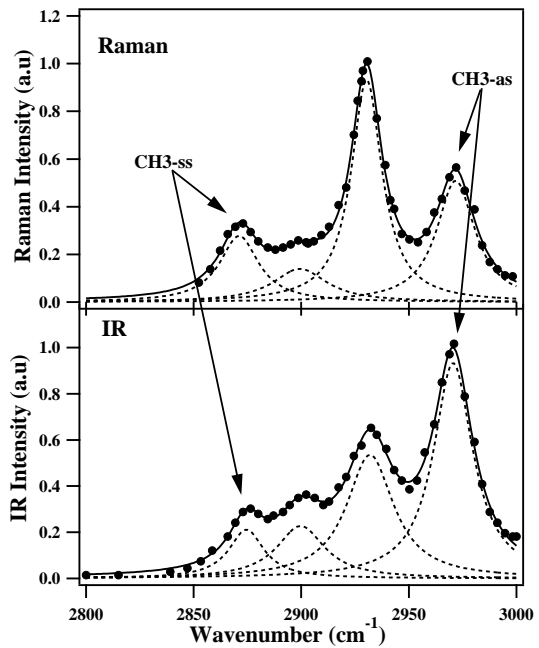}
\caption{Raman and IR spectra (solid dots, normalized to the
strongest peak in each spectra) of pure liquid CH$_{3}$CD$_{2}$OH at
300K, redrawn from previous literature.\cite{PerchardJChimP65p1856}
The dashed lines are fittings with Gaussian lineshape. The base
lines are determined by the best Gaussian group fittings. The area
of each Gaussian curve is integrated and normalized as the
corresponding peak intensity listed in Table
\ref{Ethanoltable}.\label{EthanolRamanIR}}
\end{figure}
\end{center}

The Raman and IR spectra of 12 kind of deuterated ethanol molecule
were systematically studies by Perchard and Josien in
1968.\cite{PerchardJChimP65p1856} Fig.\ref{EthanolRamanIR} shows the
Raman (measured with polarized light) and IR spectra of pure liquid
CH$_{3}$CD$_{2}$OH at 300K reproduced from Perchard and Josien's
paper. With the reproduced spectra, we performed band fitting and
listed the integrated band intensities in Table \ref{Ethanoltable}.
therefore, the intensity ratios between the CH$_{3}$-ss and
CH$_{3}$-as of the Raman and IR spectra are 0.36/0.67=0.54, and
0.20/1.00=0.20, respectively. Comparing with the values for
CH$_{3}$OH above, it is clear that the relative CH$_{3}$-as band
Raman and IR intensities are many times stronger for
CH$_{3}$CD$_{2}$OH than those for CH$_{3}$OH.

\vspace{0.3cm}

\begin{center}
\begin{table}
\caption{Integrated Raman and IR band intensities of liquid
CH$_{3}$CD$_{2}$OH at 300K. All values are normalized to the
strongest peak in the Raman and IR spectra,
respectively.\label{Ethanoltable}}
\begin{tabular}{llccc}
\hline  & Band & $\nu(cm^{-1})$ & FWHH & Intensity \\
\hline
Raman & CH$_{3}$-ss &2871 & 22 & 0.36 \\
 & CH$_{3}$-Combi &2899 & 30 & 0.25 \\
 & CH$_{3}$-Fermi &2930 & 17 & 1.00 \\
 & CH$_{3}$-as &2972 & 22 & 0.67 \\
\hline
IR & CH$_{3}$-ss &2875 & 20 & 0.20 \\
 & CH$_{3}$-Combi &2900 & 27 & 0.29 \\
 & CH$_{3}$-Fermi &2932 & 27 & 0.69 \\
 & CH$_{3}$-as &2971 & 22 & 1.00 \\
\hline
\end{tabular}
\end{table}
\end{center}

Now the empirical correction for the bond additivity model can be
readily made with Eq.\ref{IR}, Eq.\ref{BAM3}, and Eq.\ref{BetaRatio}
as the following.

For CH$_{3}$OH, with $r=0.28$, $\omega_{sym}=2836cm^{-1}$ and
$\omega_{asym}=2987cm^{-1}$, we have,

\begin{eqnarray}
\frac{\alpha'_{cc}}{\alpha'_{ac}}=1.9\mbox{ ;} \qquad
\frac{\mu'_c}{\mu'_a}=1.9\mbox{ ;} \qquad
\frac{\beta_{ccc}}{\beta_{aca}}=3.8
\end{eqnarray}

For CH$_{3}$CD$_{2}$OH, with $r=0.026$, $\omega_{sym}=2871cm^{-1}$
and $\omega_{asym}=2972cm^{-1}$, we have,

\begin{eqnarray}
\frac{\alpha'_{cc}}{\alpha'_{ac}}=0.34\mbox{ ;} \qquad
\frac{\mu'_c}{\mu'_a}=0.89\mbox{ ;} \qquad
\frac{\beta_{ccc}}{\beta_{aca}}=0.31
\end{eqnarray}

\subsection{Discussion}

Table \ref{Summarytable} summarizes the values of $R$, $r$,
$\beta_{ccc}/\beta_{aca}$, $\alpha'_{cc}/\alpha'_{ac}$,
$\mu'_{c}/\mu'_{a}$ in the uncorrected and corrected bond additivity
model. It is easy to see that the biggest correction is the
$\mu'_{c}/\mu'_{a}$ value for CH$_{3}$OH. According to the simple
bond moment theory, $\mu'_{c}/\mu'_{a}=0.48$ for both CH$_{3}$OH and
CH$_{3}$CD$_{2}$OH, i.e. the IR intensity for the CH$_{3}$-as mode
should be about 2(1/0.48)$^{2}=8.7$ times of that for the
CH$_{3}$-ss mode. However, experimentally they are about 1.2 and 5
for CH$_{3}$OH and CH$_{3}$CD$_{2}$OH, respectively. Such big
difference between these two molecules can be attributed to the fact
that there is much stronger influence on the CH$_{3}$ group
stretching vibrations by the O-H group in CH$_{3}$OH than in
CH$_{3}$CD$_{2}$OH.

It is interesting to see that $r=0.026$ for CH$_{3}$CD$_{2}$OH,
while $r=0.28$ for CH$_{3}$OH. According to the definition of $r$ in
Eq.\ref{BondAlpha}, $r=0$ means that each bond can be pictured as a
perfect rod. Therefore, the value of $r$ may be used to correlate
the coupling between the different bonds in the same group, and the
perturbation of the bonds in the group by the connecting molecular
groups. The CH$_{3}$ group in CH$_{3}$OH is likely to be much more
strongly perturbed by the directly connecting O-H group, than the
CH$_{3}$ group in CH$_{3}$CD$_{2}$OH by the directly connecting
$CD_{2}$ and the indirectly connecting O-H groups. The same effect
as in CH$_{3}$OH can also be expected for acetone and acetonitrile
molecules, where the CH$_{3}$ groups are directly connected to the
-C=O and -C$\equiv$N groups, respectively. Similarly, the longer
chain alcohols should be similar to CH$_{3}$CD$_{2}$OH. Therefore,
it is not coincidental that the asymmetric C-H stretching mode in
the SFG-VS spectra from the vapor/acetone and vapor/acetonitrile
interfaces has not been observed,\cite{YRShenJCP114p1837,
ChenHua2,ChenHua1,SomorjaiJPCB2003Acetonitrile} while for the longer
chain alcohols it has been experimentally
observed.\cite{Shen1999JPCB103p3292,ShenCPL1995Alcohol,HFWangJPCB109p14118}

\begin{center}
\begin{table}
\caption{Comparison of the bond additivity model and corrected bond
additivity model values.\label{Summarytable}}
\begin{tabular}{ccccc}
\hline& \multicolumn{2}{c}{CH$_{3}$OH} & \multicolumn{2}{c}{CH$_{3}$CD$_{2}$OH}\\
& BAM & Corrected & BAM & Corrected \\
 \hline
$R=\frac{\beta_{aac}}{\beta_{ccc}}$ & 1.7 & 1.7 & 3.4 & 3.4 \\
r & 0.28 & 0.28 & 0.026 & 0.026 \\
$\frac{\beta_{ccc}}{\beta_{aca}}$ & 1.1 & 3.8 & 0.30 & 0.31 \\
$\alpha'_{cc}/\alpha'_{ac}$ & 2.2 & 1.9 & 0.60 & 0.34 \\
$\mu'_{c}/\mu'_{a}$ & 0.48 & 1.9 & 0.48 & 0.89 \\
\hline
\end{tabular}
\end{table}
\end{center}

If we put $I_{sys}^{Raman}/I_{asym}^{Raman}=5.0$ for CH$_{3}$OH into
Eq.\ref{Ratio-r}, we have $r=0.24$ for $0\leq r\leq 1$; while if we
put $I_{sys}^{Raman}/I_{asym}^{Raman}=0.54$ for CH$_{3}$CD$_{2}$OH
into Eq.\ref{Ratio-r}, there is no solution of $r$ for $0\leq r\leq
1$ at all. This clearly shows that Eq.\ref{rho-r} and
Eq.\ref{Ratio-r} do not generally give consistent $r$ values, and
certainly they can not be held true at the same time.

Now, with the corrected $\beta_{ccc}/\beta_{aca}$ value for
CH$_{3}$OH, the fact of missing CH$_{3}$-as mode in SFG-VS from
vapor/CH$_{3}$OH interface can be readily addressed. From the
calculated intensities in Fig.\ref{MethanolSFG} with
Vis=62$^{\circ}$ and IR=53$^{\circ}$ experimental configuration,
$I_{CH_{3}-as}^{sps}/I_{CH_{3}-ss}^{ssp}\approx 1/2.5$ with the
uncorrected bond additivity model value
$\beta_{ccc}/\beta_{aca}=1.1$, and with the known orientation angle
$\theta\approx 0$ from previous
studies.\cite{Lurong1,GanweiCPLNull,ChenHua3} Now, with the
corrected value $\beta_{ccc}/\beta_{aca}=3.8$, the
$I_{CH_{3}-as}^{sps}/I_{CH_{3}-ss}^{ssp}$ ratio should be
(3.8/1.1)$^{2}=11.9$ times smaller, i.e.,
$I_{CH_{3}-as}^{sps}/I_{CH_{3}-ss}^{ssp}\approx 1/30$. It is easy to
see that such a small $I_{CH_{3}-as}^{sps}$ comparing to
$I_{CH_{3}-ss}^{ssp}$ is below the noise level and its spectral
feature can not be observed in the SFG-VS spectra. Convincingly,
using above approach, inspection of the IR and Raman spectra of
liquid acetone and acetonitrile also indicated that the CH$_{3}$-as
vs. CH$_{3}$-ss intensity ratio values are significantly smaller
than the values calculated from the simple bond additivity model for
their SFG-VS spectra from vapor/liquid
interfaces.\cite{YRShenJCP114p1837,ChenHua2,ChenHua1,SomorjaiJPCB2003Acetonitrile}

It is interesting to see that the correction for the
$\beta_{ccc}/\beta_{aca}$ value is very small for
CH$_{3}$CD$_{2}$OH, even though the corrections for
$\alpha'_{cc}/\alpha'_{ac}$ and $\mu'_{c}/\mu'_{a}$ are not so small
separately. These two corrections happened to cancel each other when
used to calculate $\beta_{ccc}/\beta_{aca}$ with Eq.\ref{BetaRatio}.
Using this ratio, the SFG-VS spectra of CH$_{3}$CD$_{2}$OH can be
well interpreted just as with the uncorrected bond additivity
model.\cite{HongfeiIRPCreview} However, it is to be noted that the
success of the former is with firm empirical basis, and the success
of the latter is nevertheless coincidental. We have performed
measurements on the SFG-VS spectra in different incident angles for
the vapor/ethanol interface, and analysis show that with the
$\beta_{i'j'k'}$ tensor ratios obtained here, all spectra as well as
the spectra interference effects, can be quantitatively interpreted,
and detailed molecular orientation analysis is also
possible.\cite{GanWeiSymmetryAnalysis}

According to above discussion, the success of the corrected bond
additivity model is evident. This success indicates that the failure
of the bond additivity model discussed in Section IIB can be
quantitatively corrected with the empirical approach as demonstrated
here. However, we have to caution that this success is purely
empirical, and may not be as general as it seems before further
examinations being made. Even though it is undoubtedly an
advancement from the simple bond additivity model, there are reasons
this empirical approach may fail.

Because the correction for the bond additivity model used the Raman
and IR spectra in the liquid phase, the success of this correction
undoubtedly indicates that the CH$_{3}$ group of these molecules
under studying are not significantly different as in the liquid bulk
phases or at the interfaces. This may only be true for the generally
`rigid' C-H stretching vibrations which is usually not greatly
perturbed and the assumption of local motion is generally valid. But
this may not be true for the less `rigid' molecular groups. Based on
the successful treatment with the CH$_{3}$ groups, we may assume
that the same treatment on the similarly `rigid' NH$_{3}$ and
CH$_{2}$ groups is also going to work. If this approach fails to
quantitatively interpret the SFG-VS spectra for a particular
molecular group, one may try to consider the possibility that this
group might have been very differently perturbed in the bulk phase
from at the interface.

One clear advantage of this empirical approach is that the problem
of the Fermi resonance and combinational mode couplings is
automatically addressed through the empirical correction. As we have
known, to assess the effects of such accidental resonances and
couplings can be a very difficult problem in \textit{ab initio}
calculations. Therefore, the empirical treatment discussed here may
provide a way to quantitatively test the whole molecule
approach.\cite{RichmondJPCBp16846} For example, one direct may is to
try to use the whole molecule approach to calculate the
polarizability tensors of the CH$_{3}$ group in the CH$_{3}$OH and
CH$_{3}$CD$_{2}$OH, and compare the results directly with the IR,
Raman and SFG-VS data. Provided with successful applications to a
series of clearly studied molecular systems, one may expect that in
the future \textit{ab initio} computation can be generally applied
for quantitative interpretation of nonlinear spectroscopy.

\section{Conclusion}

In summary, we discussed a total empirical approach with symmetry
analysis to quantitatively interpret observed SFG-VS spectra from
molecular interfaces. In such an approach, the Raman and IR tensor
ratios are directly calculated from Raman and IR spectral
measurements using a certain correction procedure to the bond
additivity model. This approach is tested with the SFG-VS spectra
from vapor/CH$_{3}$OH and vapor/CH$_{3}$CD$_{2}$OH interfaces.

There have been few attempts to use Raman and IR spectra in the
condensed phase to quantitatively interpret SFG-VS spectra from
molecular interfaces in detail. It has been demonstrated that the
bond additivity model has been successful in quantitative
interpretation of SFG-VS of the CH$_{3}$ and CH$_{2}$ as well as
H$_{2}$O groups for their ss modes,\cite{HongfeiIRPCreview} but it
has not been as successful with the as of these molecules or
molecular groups. Through analysis of the problem we understood the
empirical basis for the successes of the bond additivity model. So
we proposed to treat the failures of the bond additivity model with
a complete empirical approach. In this report we successfully
demonstrated with two examples that the empirical corrections to the
bond additivity model can successfully interpret what the simple
bond additivity model failed. This development can become a new
addition to the existing methodologies in quantitative SFG-VS
studies.\cite{HongfeiIRPCreview} We believe that this empirical
approach can provide new understanding of the effectiveness and
limitations of the bond additivity model, and this report also
provided a practical roadmap for its application in SFG-VS studies
of molecular interfaces.

\vspace{0.5cm}

\noindent \textbf{Acknowledgment.} HFW thanks supports by the
Natural Science Foundation of China (NSFC No.20274055, No.20425309,
No.20573117). ZFC thanks support by the Key Project of Ministry of
Education of China (MOE No.204065).

\section*{APPENDIX: Terms and expressions for C$_{2v}$ group corrections}

With the empirical approach derived and tested for the C$_{3v}$
groups, similar approach can be taken for the C$_{2v}$ groups. For
them, bond additivity model has to be used to obtain necessary
$\beta_{i'j'k'}$ tensor ratios for quantitative
calculation.\cite{YRShenPRE62p5160,DuQuanPRL1993,HFWangJPCB108p7297,HFWangJPCB109p14118}

If the fixed coordinates system of the AB$_{2}$ group with C$_{2v}$
symmetry is defined with $c$ axis as the symmetry axis, $ac$ plane
as the A-B-A molecular plane, and $b$ is perpendicular to the
molecular plane, then there are two A-B stretching vibrational mode
with $A_{1}$ mode as the symmetric stretching (ss) vibrational mode,
and $B_{1}$ mode the asymmetric (as)
mode.\cite{NafieJCP1972,Longbook,HongfeiIRPCreview} The non-zero
elements for the stretching vibrational modes are as the followings.

For the $A_{1}$ (symmetric) mode:
\begin{eqnarray}
\mbox{SFG-VS \ \ } &&
\beta_{aac}=R_{a}\beta_{ccc}\mbox{ ;} \qquad \beta_{bbc}=R_{b}\beta_{ccc}\label{SFGssC2v}\\
\mbox{Raman: \ \ } &&
\alpha'_{aa}=R_{a}\alpha'_{cc} \mbox{  ;} \qquad \alpha'_{bb}=R_{b}\alpha'_{cc}\label{RamanssC2v} \\
\mbox{IR: \ \ } && \mu'_{c}
\end{eqnarray}

\noindent where R$_{a}$ and R$_{b}$ are the two $\beta_{i'j'k'}$
elements ratios.

For the $B_{1}$ (asymmetric) mode:
\begin{eqnarray}
\mbox{SFG-VS \ \ } &&
\beta_{aca}=\beta_{caa}\label{SFGasC2v}\\
\mbox{Raman: \ \ } && \alpha'_{ac}=\alpha'_{ca}\label{RamanasC2v}\\
\mbox{IR: \ \ } && \mu'_{a}
\end{eqnarray}

Therefore, Eq.\ref{IR} for IR intensity of the $C_{2v}$ group
becomes,

\begin{eqnarray}
\frac{I_{sym}^{IR}}{I_{asym}^{IR}}=\left(\frac{\mu'_{c}}{\mu'_{a}}\right)^2\label{IRC2v}
\end{eqnarray}

\noindent Eq.\ref{RamanDPRatio}-Eq.\ref{gamma} are all generally
valid for Raman
measurement,\cite{NafieJCP1972,Longbook,HongfeiIRPCreview} and
Eq.\ref{rhodefinition} alone cannot be used to determine both
R$_{a}$ and R$_{b}$ values. Therefore, bond additivity model has to
be employed.

For the single A-B bond treated with $C_{\infty v}$ symmetry,
Eq.\ref{BondAlpha} is generally assumed valid, and the definition of
$a_{0}$ and $m_{0}$ of the single bond is the same as defines above.
For the C$_{2v}$ group AB$_{2}$ with the A-B-A angle as $\tau$, we
have the following:

\begin{eqnarray}
&&G_{A_{1}}=\frac{1+\cos{\tau}}{M_A}+\frac{1}{M_B}\nonumber\\
&&G_{B_{1}}=\frac{1-\cos{\tau}}{M_A}+\frac{1}{M_B}\nonumber
\end{eqnarray}

\noindent in which $M_{A}$ and $M_{B}$ are the atomic masses of the
atom $A$ and $B$; $G_{A_{1}}$ and $G_{B_{1}}$ are the inverse
reduced masses of the $A_{1}$ and $B_{1}$ normal modes,
respectively, with the normal mode coordinates defined
as,\cite{HiroseJCP96p997,HiroseJPC97p10064,WubaohuaThesis}

\begin{eqnarray}
&&Q_{A_{1}}=(\Delta r_{1}+\Delta r_{2})/(2G_{A_{1}})^{1/2}\nonumber\\
&&Q_{B_{1}}=(\Delta r_{1}-\Delta r_{2})/(2G_{B_{1}})^{1/2}\nonumber\\
\end{eqnarray}

Then the bond additivity model gives the following expressions of
the non-zero $\alpha'_{i'j'}$ and $\mu'_{k'}$ tensors.

For the $A_{1}$ (symmetric) mode:
\begin{eqnarray}
&&\alpha'_{aa}=a_{0}[\sin^{2}(\tau/2)+r\cos^{2}(\tau/2)]{(2G_{A_{1}})}^\frac{1}{2}\nonumber\\
&&\alpha'_{bb}=a_{0}r{(2G_{A_{1}})}^\frac{1}{2}\nonumber\\
&&\alpha'_{cc}=a_{0}[r\sin^{2}(\tau/2)+\cos^2{\tau/2}]{(2G_{A_{1}})}^\frac{1}{2}\nonumber\\
&&\mu'_{c}=m_{0}\cos(\tau/2){(2G_{A_{1}})}^\frac{1}{2}\nonumber
\end{eqnarray}

For the $B_{1}$ (asymmetric) mode:
\begin{eqnarray}
&&\alpha'_{ac}=\alpha'_{ca}=\frac{1}{2}a_{0}[(1-r)\sin{\tau}]{(2G_{B_{1}})}^\frac{1}{2}\nonumber\\
&&\mu'_{a}=m_{0}\sin(\tau/2){(2G_{B_{1}})}^\frac{1}{2}\nonumber
\end{eqnarray}

\noindent Therefore, the bond additivity gives the following
$\beta_{i'j'k'}$ tensor ratios.\cite{HongfeiIRPCreview}

\begin{eqnarray}
&&R_{a}=\frac{\beta_{aac}}{\beta_{ccc}}=
\frac{1+r-(1-r)\cos\tau}{1+r+(1-r)\cos\tau}\nonumber\\
&&R_{b}=\frac{\beta_{bbc}}{\beta_{ccc}}=\frac{2r}{1+r+(1-r)\cos\tau}\nonumber\\
&&R_{c/a}=\frac{\beta_{ccc}}{\beta_{aca}}=\frac{\beta_{ccc}}{\beta_{caa}}\nonumber\\
&&=\frac{[(1+r)+(1-r)\cos\tau]\cos(\tau/2)}{(1-r)\sin\tau\sin(\tau/2)}\frac{G_{A_{1}}\omega_{B_{1}}}{G_{B_{1}}\omega_{A_{1}}}
\label{RC2vdefinition}
\end{eqnarray}

\noindent with the $r$ values calculated from the experimentally
obtained Raman depolarization ratio $\rho$ of the ss mode of the
AB$_{2}$ group using the following equation.

\begin{eqnarray}
\rho&=&\frac{3}{4+20(1+2r)^2/[(1-r)^2(1+3\cos^2\tau)]}\label{rCH2definition}
\end{eqnarray}

Following the example of the C$_{3v}$ group above, the procedure to
make correction for the bond additivity model is as the following.
First to use Eq.\ref{rCH2definition} to obtain $r$ value. Then the
$R_{a}$ and $R_{b}$ values can be obtained using this $r$ value. In
order to calculate $R_{c/a}$, one need to use Eq.\ref{BetaRatio}
with the empirically corrected $\alpha'_{cc}/\alpha'_{ac}$ and
$\mu'_{c}/\mu'_{a}$ values. The $\mu'_{c}/\mu'_{a}$ value can be
obtained using Eq.\ref{IRC2v}. Then the $\alpha'_{cc}/\alpha'_{ac}$
value can be obtained by put the $r$ value and the bond additivity
model expressions for $\alpha'_{i'j'}$ tensors above into
Eq.\ref{RamanRatio} to derive a equation similar to
Eq.\ref{RatioCorrection}.

This correction procedure is certainly worth to be tested for
interpretation of SFG-VS data of typical C$_{2v}$ molecules or
molecular groups at interfaces. For example, at the air/water and
organic/water interfaces, so far the asymmetric stretching mode of
the C$_{2v}$ water molecule has not been clearly identified
yet.\cite{DuQuanPRL1993,HongfeiCJCPWater,HongfeiJCPWaterLong}
Another example is that the intensity ratio between CH$_{2}$-ss and
CH$_{2}$-as has been widely used to interpret CH$_{2}$ group
orientational changes at molecular and polymer
interfaces.\cite{SomorjaiJPCB106p5212}

\end{document}